\documentclass[12pt]{article}
\usepackage[dvips]{graphics}

\catcode`\@=11
\def\marginnote#1{}
\newcount\hour
\newcount\minute
\newtoks\amorpm
\hour=\time\divide\hour by60
\minute=\time{\multiply\hour by60 \global\advance\minute
by-\hour}\edef\standardtime{{\ifnum\hour<12
\global\amorpm={am}%
        \else\global\amorpm={pm}\advance\hour by-12 \fi
        \ifnum\hour=0 \hour=12 \fi
        \number\hour:\ifnum\minute<10
0\fi\number\minute\the\amorpm}}
\edef\militarytime{\number\hour:\ifnum\minute<10
0\fi\number\minute}

\def\draftlabel#1{{\@bsphack\if@filesw {\let\thepage\relax
   \xdef\@gtempa{\write\@auxout{\string
      \newlabel{#1}{{\@currentlabel}{\thepage}}}}}\@gtempa
   \if@nobreak \ifvmode\nobreak\fi\fi\fi\@esphack}
        \gdef\@eqnlabel{#1}}
\def\@eqnlabel{}
\def\@vacuum{}
\def\draftmarginnote#1{\marginpar{\raggedright\scriptsize\tt#1}}
\def\draft{\oddsidemargin -.5truein
        \def\@oddfoot{\sl preliminary draft \hfil
        \rm\thepage\hfil\sl\today\quad\militarytime}
        \let\@evenfoot\@oddfoot \overfullrule 3pt
        \let\label=\draftlabel
        \let\marginnote=\draftmarginnote

\def\@eqnnum{(\theequation)\rlap{\kern\marginparsep\tt\@eqnlabel}%
\global\let\@eqnlabel\@vacuum}  }

\def\numberbysection{\@addtoreset{equation}{section}
        \def\theequation{\thesection.\arabic{equation}}}

\def\underline#1{\relax\ifmmode\@@underline#1\else
 $\@@underline{\hbox{#1}}$\relax\fi}

\catcode`@=12
\relax

\numberbysection
\def\demi{{1\over 2}}
\def\r2{\sqrt{2}}
\def\beq{\begin{equation}}
\def\eeq{\end{equation}}
\def\bea{\begin{eqnarray}}
\def\eea{\end{eqnarray}}
\def\nnn{\nonumber \\}

\def\Gcc#1 {{\cal G}^{[ #1 ]}_\|}
\def\Gcp#1 {{\cal G}^{[ #1 ]}_\bot}
\def\Cco#1 {{\cal C}^{[ #1 ]} }

\def\1{{{\dot 1}}}
\def\2{{{\dot 2}}}
\def\3{{{\dot 3}}}
\def\4{{{\dot 4}}}
\def\5{{{\dot 5}}}
\def\6{{{\dot 6}}}
\def\7{{{\dot 7}}}
\def\8{{{\dot 8}}}
\def\fin{\end{document}}

\def\d(#1,#2){\delta_{#1,\, #2}}
\begin{document} 
\begin{titlepage}
\begin{flushright}

LPTENS--97/64, \\
hep-th/9810172, \\
October  1998
\end{flushright}

\vglue 2.5  true cm
\begin{center}
{\large
T Duality Between  Perturbative Characters\\  
\medskip
of $E_8\otimes E_8$ and $SO(32)$ Heterotic Strings 
\\ \medskip Compactified On A Circle\footnote
{Work supported in part by the TMR contract FMRX-CT96-0012i.} }  \\
\vglue 1.5 true cm
{\bf Jean-Loup~GERVAIS}\\
\medskip
{\footnotesize Laboratoire de Physique Th\'eorique de
l'\'Ecole Normale Sup\'erieure\footnote{Unit\'e Propre du
Centre National de la Recherche Scientifique,
associ\'ee \`a l'\'Ecole Normale Sup\'erieure et \`a
l'Universit\'e
de Paris-Sud.},\\
24 rue Lhomond, 75231 Paris CEDEX 05, ~France}.
\end{center}
\vfill
\begin{abstract}
\baselineskip .4 true cm
\noindent
{\footnotesize
Characters of $E_8\otimes E_8$ and $SO(32)$ heterotic strings involving the full internal symmetry
Cartan subalgebra generators are defined after circle compactification so that they are T dual. The
novel point, as compared with an earlier study of the type II case, is the appearence of Wilson
lines. Using  $SO(17,1)$ transformations between the weight lattices reveals  the existence of an
intermediate theory where T duality transformations are disentangled from the internal symmetry.  
This intermediate theory corresponds to a sort of twisted compactification of a novel type. Its
modular invariance follows from an interesting interplay between three representations of the 
modular group.
 }
\end{abstract}
\vfill
\end{titlepage}
\tableofcontents
\section{Introduction} 
In general, string theories  are invariant under a certain global
 symmetry group $G$, of rank $R$,  that
commutes with the Virasoro algebra. It is thus natural to consider 
the generalisation of the usual partition torus functions which mathematically is  
the character\footnote{Of course, we are really dealing with Kac-Moody type characters. This point is
not central here, and will not be considered.} of the representation of 
$G\otimes $ Vir.,  span by the free string states;    and is    of the type 
$$
\chi(\tau \,|\, v_1, \ldots, v_R)={\rm Tr}\left\{
e^{2i\pi \tau \left(L_0-a\right)} e^{2i\pi \tau^* \left(L^*_0-\widetilde a\right)}\prod_{j=1}^R
e^{2i\pi v_jH_j}
\right\}.   
$$
The trace is taken over all the physical string states, and $\exp \left(2i\pi v_jH_j\right)$
form a maximal set of commuting  generators of  $G$. 
The usual partition function 
$P(\tau)={\rm Tr}\left \{ e^{2i\pi \tau \left(L_0-a\right)} e^{2i\pi \tau^*
\left(L^*_0-\widetilde a\right)}\right\}$ is  clearly recovered when the $v$'s vanish. The elliptic
genus, which has been much studied (see, e.g. ref.\cite{DMVV}) corresponds to the case
 where $G$ is chosen to be generated by the left- and right-moving
fermionic number operators.  
On the other hand, 
$G$ contains the group of orthogonal transverse rotations of the string components in the
uncompactified directions, and the associated Lie group characters were studied some time
ago\cite{anciens1}, \cite{anciens2}  as  book
keeping devices for the representation content of string theories.  The recent developments give us
a strong motivation to study their duality and modular  properties.  These aspects  left
over in previous discussions, were recently studied in details in ref.\cite{CG} for perturbative
type II strings. The present article deals with  heterotic perturbative string
characters along the same line.  Here it is of course natural to extend
$G$  to involve the internal symmetry group, that is,  either $SO(32)$ or $E_8\otimes E_8$, as was
done originally\cite{anciens2}. 
Without any compactification, the  object under study is of the type 
\beq
\chi(\tau \,|\, \vec v,\,  \vec \xi)={\rm Tr}\left\{
e^{2i\pi \tau \left(L_0-a\right)} e^{2i\pi \tau^* \left(L^*_0-\widetilde a\right)}\prod_{j=1}^4
e^{2i\pi v_j{\cal  H}_j} \prod_{j=1}^{16}
e^{2i\pi \xi_j{\cal  A}_j}, 
\right\},   
\label{chigen}
\eeq
where $\left\{{\cal  H}_1, \ldots {\cal  H}_4\right\}$ are four commuting transverse-space rotation
generators, and $\left\{{\cal  A}_1, \ldots {\cal  A}_{16}\right\}$ are sixteen commuting 
 generators of  the  internal symmetry group. 
 
  The novel point of the present study, as compared to ref.\cite{CG}, 
is that duality between the two compactified heterotic theories requires\cite{G} that one breaks
their internal symmetry group, so that the ideas just summarized do not apply strictly speaking. 
The point of this paper
is to show how to define, nevertheless, the  characters involving sources for the
full Cartan subalgebras of each  internal symmetry group  so 
that they are T dual. 
The corresponding  equality generalises the well known equality between partition functions (see e.g.
\cite{GSW}) which is of course recovered when all the  $v$'s and
$\xi$'s  vanish. The basic mathematical tool to derive  T duality of the circle
compactified heterotic strings is the  
$SO(17,1)$ transformation between the two   embeddings of   $E_8\otimes E_8$ and spin$(32)/Z_2$ zero
mode lattices into the same Lorentzian lattice $\Pi^{17,1}$.  It may be factored into three
transformations.  This shows the existence of an intermediate theory where the internal symmetry
group is broken down to
$SO(16)\otimes SO(16)$ by a sort of twisted compactification of a novel type. The character is then 
straightforwardly defined in the intermediate theory since this unbroken gauge group has the same
Cartan subalgebra as the broken ones.  This intermediate theory is such that it
interpolates between the two uncompactified heterotic strings which are recovered for very large and
very small compactification radius, respectively. The  modular
invariance of the character follows from an interesting interplay
between three representations of the  modular group. Finally, the characters of the original 
$E_8\otimes E_8$ and spin$(32)/Z_2$ heterotic theories are defined in a natural way using the 
$SO(17,1)$ transformations which map their weight lattices to the one of the intermediate theory.

\section{The uncompactified characters of  heterotic strings }
This case was already discussed in ref.\cite{anciens2}. 
In the same way as for type II superstring, we  introduce modified  characters, by also summing over
the total momentum, following the line of ref.\cite{CG}.  We shall be brief and refer to the last
two references for details. This section is a sort of warming up for the coming discussion of the
compactified case.  The characters may be factorised under the form
\beq
\chi(\tau \,|\, \vec v,\, \vec \xi)=\chi_0(\vec v)\chi_L(\tau \,|\, \vec v )
\chi_R(\tau \,|\, \vec v ,\, \vec \xi), 
\label{chifac}
\eeq
where $\chi_0$ comes from the summation over the transverse components of the total momentum, 
and $\chi_L$,$\chi_R$ are, respectively,  the contributions of the left and right transverse
string modes. We select  the right movers to be the standard compactified bosonic modes. The
$v_i$'s  are the same, for  left and right  wordsheet fields since they   transform at the same
time under space rotations. The  first term  in the  equation above was computed in ref.\cite{CG}.
The result reads\footnote{We drop overall constant factors throughout.}
\beq
\chi_0(\vec v)= \prod_{k=1}^4 {1\over \sin^2\left(\pi v_k \right)}. 
\label{chi0}
\eeq 
The second term  is the same as for type II. In the notations of  ref.\cite{CG}), we choose  the
chirality sector  GS$_-$
  for
definiteness. The left factor  of the character is thus  
\beq 
\chi_{L}(\tau \,|\, \vec v )=\left(\chi_{{\rm  GS}_-}(\tau\,|\, \vec v)\right)^*\equiv 
\left( \prod_{k=1}^4\sin\left(\pi v_k \right)
{\Theta^{(8)} _1(\vec y\,|\,\tau )\over\Theta^{(8)}_1(\vec v\,|\,\tau )}\right)^*
\label{chiL}
\eeq
The notations are the same as in ref.\cite{CG} apart from minor details.  
   The function $\Theta^{(2N)}$ are defined in
general by  
\beq
\Theta^{(2N)}_i(\vec x\,|\, \tau)=
\prod_{\mu=1}^{N} \theta_i( x_\mu |\, \tau)
\label{Thetadef}
\eeq
where $\theta_i$ are the usual theta functions with Bateman's conventions. 
The vector $\vec y$ is obtained from $\vec v$ by the  orthogonal transformation
associated to one of the $SO(8)$ triality transformations (see ref.\cite{CG}). Accordingly  we have
the  Jacobi identity
$$
\Theta^{(8)}_1(\vec y\,|\,\tau)=\demi\left(
\Theta^{(8)}_1(\vec v\,|\,\tau)-\Theta^{(8)}_2(\vec v\,|\,\tau)
+\Theta^{(8)}_3(\vec v\,|\,\tau)-\Theta^{(8)}_4(\vec v\,|\,\tau)\right). 
$$
In order to ensure modular invariance, the character is defined by taking the supertrace.   
  The  last factor of Eq.\ref{chifac} involves the internal symmetry group, and thus 
should be discussed  separately for the two heterotic strings. 
\subsection{The $SO(32)$ case. }
The right character 	may be factorised as follows 
\beq
\chi_R(\tau \,|\, \vec v,\,  \vec \xi)=e^{-2i\pi \tau} 
{\prod_{k} \sin\left(\pi v_k \right)\over \Theta^{(8)}_1(\vec v\,|\,\tau )}
\chi^{{\rm O }}(\tau \,|\, \vec \xi), 
\label{chiR16fac}
\eeq
where the explicit factors are the contribution of  the space bosonic components.  
The calculation of  $\chi^{{\rm O }}$  is similar to the discussion  of the 
NSR characters recently  carried out in ref.\cite{CG}.  Some standard features of the world-sheet
properties of heterotic strings  are summarized in appendix A. Altogether, the following is
straightforwardly derived.
 
\paragraph{The P-right sector}
As recalled in appendix A, the  intercept, is  $ a_P=-1$ giving an  
overall factor $e^{2i\pi \tau}$. One finds  
$$
\chi^{{\rm O }}(\tau \,|\, \vec \xi)=e^{4i\pi \tau} \demi 
\left\{  \left[\prod_{n,\,\nu=1}^{16}
(1+e^{2in\pi \tau}e^{2i\pi \xi_\nu} )(1+e^{2in\pi \tau}e^{-2i\pi \xi_\nu} )\right.\right.
$$
$$
\left.
+ \prod_{n,\,\nu=1}^{16}
(1-e^{2in\pi \tau}e^{2i\pi \xi_\nu} )(1-e^{2in\pi \tau}e^{-2i\pi \xi_\nu} )\right]
\chi_{(0,\cdots ,0,1)}^{O(32)}(\vec \xi)+
$$
$$
\left[\prod_{n,\,\nu=1}^{16}
(1+e^{2in\pi \tau}e^{2i\pi \xi_\nu} )(1+e^{2in\pi \tau}e^{-2i\pi \xi_\nu} )\right.
$$
$$
\left.\left.
- \prod_{n,\,\nu=1}^{16}
(1-e^{2in\pi \tau}e^{2i\pi \xi_\nu} )(1-e^{2in\pi \tau}e^{-2i\pi \xi_\nu} )\right]
\chi_{(0,\cdots ,1,0)}^{O(32)}(\vec \xi)\right\}. 
$$ 
The symbols $\chi_{(0,\cdots ,0,1)}^{O(32)}$, and $\chi_{(0,\cdots ,1,0)}^{O(32)}$ denote the 
irreducible characters of the two fundamental spinor representations. 
\paragraph{A-right sector}
Now  the intercept $ a_A=1$, and one finds  
$$
\chi^{{\rm O }}_A(\tau\,| \vec \xi)= 
\demi\left\{ \prod_{r=1/2}^\infty \prod_{\mu=1}^{16} 
(1+e^{2i\pi r\tau}e^{2i\pi \xi_\mu} )((1+e^{2i\pi r\tau}e^{-2i\pi \xi_\mu} )\right.
$$
$$\left. +
\prod_{r=1/2}^\infty \prod_{\mu=1}^{16} (1-e^{2i\pi r\tau}e^{2i\pi \xi_\mu} )
((1-e^{2ir\pi \tau} e^{-2i\pi \xi_\mu} )\right\}
$$
Next we re-express these characters in terms of Theta functions. For this 
one has  to simplify the expressions by getting rid of $\chi_{(0,\cdots ,0,1)}^{O(32)}$ and 
$\chi_{(0,\cdots ,1,0)}^{O(32)}$. In the same way as for the  $O(8)$ case discussed on
ref.\cite{CG}, one may verify that 
$$
\chi_{(0,\cdots ,1,0)}^{O(32)}(\vec \xi)
=\sum_{{\epsilon_1,\ldots \epsilon_{16}=\pm 1\atop 
{\rm odd \>  number }=1}} \prod_{\nu=1}^{16} e^{i\pi \xi_\nu\epsilon_\nu}
$$
$$
\chi_{(0,\cdots ,0,1)}^{O(32)}(\vec \xi)
=\sum_{{\epsilon_1,\ldots \epsilon_{16}=\pm 1\atop 
{\rm even \>  number }=1}} \prod_{\nu=1}^{16} e^{i\pi \xi_\nu\epsilon_\nu}
$$
A straightforward calculation gives 
\beq
\chi_P(\tau\,|\vec q,\, \vec \xi)=
f_0^{-16}\demi\left[\Theta^{(32)}_2(\vec\xi\, |\, \tau)+\Theta^{(32)}_1(\vec\xi\, |\, \tau)\right]
\label{hetP}
\eeq
\beq
\chi_A(\tau\,|\vec q,\, \vec \xi)=
f_0^{-16}\demi\left[\Theta^{(32)}_3(\vec\xi\, |\, \tau)+\Theta^{(32)}_4(\vec\xi\, |\, \tau)\right]
\label{hetA}
\eeq
where 
\beq
f_0=\prod_{n\geq 1}\left(1-e^{2i\pi n\tau}\right). 
\label{f0def}
\eeq
Next we introduce summations over the internal symmetry lattice, by 
using the series expansion of the theta functions. It is easy to derive in general 
the lattice expansion
\begin{eqnarray}
\demi \left[\Theta^{(2N)}_2(\vec \xi\, |\, \tau)+
\Theta^{(2N)}_1(\vec\xi\, |\, \tau)\right]&=&\sum_{\vec m\in \Lambda_W^{{(N)}e,h}}  
 e^{i\pi \tau \sum_\mu m_\mu^2} e^{2i\pi \sum_\mu m_\mu\xi_\mu}
\nnn
\demi \left[\Theta^{(2N)}_2(\vec \xi\, |\, \tau)-
\Theta^{(2N)}_1(\vec\xi\, |\, \tau)\right]&=&\sum_{\vec m\in \Lambda_W^{{(N)}o,h}}  
 e^{i\pi \tau \sum_\mu m_\mu^2} e^{2i\pi \sum_\mu m_\mu\xi_\mu}
\nnn
\demi \left[\Theta^{(2N)}_3(\vec \xi\, |\, \tau)+
\Theta^{(2N)}_4(\vec\xi\, |\, \tau)\right]&=&\sum_{\vec m\in\Lambda_W^{{(N)}e,i}}  
 e^{i\pi \tau \sum_\mu m_\mu^2} e^{2i\pi \sum_\mu m_\mu\xi_\mu}
\nnn
\demi \left[\Theta^{(2N)}_3(\vec \xi\, |\, \tau)-
\Theta^{(2N)}_4(\vec\xi\, |\, \tau)\right]&=&\sum_{\vec m\in\Lambda_W^{{(N)}o,i}}  
 e^{i\pi \tau \sum_\mu m_\mu^2} e^{2i\pi \sum_\mu m_\mu\xi_\mu},
\label{Theta2Npm}
\end{eqnarray} 
We will only consider the case where $N$ is multiple of $4$. The four above lattices are
equivalence classes of the weight lattice of  $O(2N)$: 
$$
\Lambda_W^{(N)}=\Lambda^{{(N)}e,h} \oplus \Lambda^{{(N)}o,h}\oplus 
\Lambda^{{(N)}e,i}\oplus \Lambda^{{(N)}o,i}
$$
The root lattice is  $\Lambda^{(N)e,i}$.   
The weight lattice $\Lambda_W^{(N)}$ is made up with the points $\sum_\mu \vec e_\mu m_\mu$ 
with $\vec e_\mu$  basis unit vectors, such that all $m_\mu$'s
are integer or half integer simultaneously. The  four
equivalence classes are conveniently  defined by further separating the points according to whether
$\sum_{\mu=1}^Nm_\mu$ is even or odd. The notation may be  summarized  as follows.   
\begin{center}
\begin{tabular}{||c|cccc||}\hline\hline 
sublattice&$\Lambda^{(N)e,i}$ &$\Lambda^{(N)o,i}$ &$\Lambda^{(N)e,h}$&$\Lambda^{(N)o,h}$ \\
$m_\mu$           & integer & integer  & half int.& half int. \\  
$\sum_\mu m_\mu $ & even  & odd &even &odd     \\ 
   \hline \hline 
\end{tabular}
\end{center} 
Altogether 
$$
\chi^{\rm O }_A(\tau\,| \vec \xi)+\chi^{\rm O }_P(\tau\,| \vec \xi)= f_0^{-16} \sum_{\vec m\in
\Gamma^{16} }  
 e^{i\pi \tau \sum_\mu (m_\mu)^2} e^{2i\pi \sum_\mu m_\mu\xi_\mu}
$$
where
\beq
\Gamma^{16}=\Lambda^{{(16)e,h}}\oplus \Lambda^{{(16)e,i}}\equiv 
\Lambda^{{(16)e,h}}\oplus \Lambda_R^{{(16)}}
\label{gamma16}
\eeq
 This lattice is of course the usual self-dual lattice (see e.g. ref.\cite{GSW} vol 1).  
let us finally put  together left-,  right-movers, and zero modes.  
One
finds, using standard identities about theta functions, 
$$
\chi^{\rm O }(\tau\,|\vec v,\, \vec \xi)=
\left({\Theta^{(8)}_1(\vec y\,|\,\tau)\over\Theta^{(8)}_1(\vec v \,|\,\tau)}\right)^*
{\left(\theta'(0|\tau)\right)^{4/3}  \over\Theta^{(8)}_1(\vec v \,|\,\tau)}
$$
\beq
\left(\theta'(0|\tau)\right)^{-16/3}\demi\left[\Theta^{(32)}_1(\vec\xi\, |\, \tau)+
\Theta^{(32)}_2(\vec\xi\, |\, \tau)+
\Theta^{(32)}_3(\vec\xi\, |\, \tau)+
\Theta^{(32)}_4(\vec\xi\, |\, \tau)\right]
\label{ch32}
\eeq
Our next point is    modular invariance.  The invariance under $\tau\to\tau+1$ is immediate.
Concerning the other basic one, that is $\tau\to -1/\tau$, one has    
$$
{\Theta^{(2N)}_i\left(\vec \xi/\tau \,|\, -1/\tau\right)\over 
\left(\theta'(0|-1/\tau)\right)^{N/3} }=\sum_j S_{ij}  
{e^{i\pi \vec \xi^2\/\tau}\Theta^{(2N)}_i\left(\vec \xi \,|\, \tau\right)\over
\left(\theta'(0|\tau)\right)^{N/3} }
\quad i=1, 2, 3, 4, 
$$
where 
\beq
S= \left(\begin{array}{cccc}1&0&0&0\\0&0&0&1\\0&0&1&0\\0&1&0&0\\ \end{array}\right). 
\label{Sdef}
\eeq
Thus 
 $$
\chi^{\rm O }(\tau\,|\vec v,\, \vec \xi)=e^{-i\pi \left(\vec \xi^2-\vec v^2\right)/\tau}
\chi^{\rm O }\left({-1\over \tau}\,\Bigl |{\vec v\over \tau },\, {\vec \xi\over \tau }\right)
$$
This is not quite modular invariant. The left over multiplicative factor is removed by using the 
simple identity
$$
{1\over \Im(\tau)}-{1\over \tau^2}{1\over \Im(-1/\tau)}={1\over \tau}. 
$$
Thus we define 
$$
\chi_{{\rm modular }}^{\rm O }=e^{i\pi \left(\vec \xi^2-\vec v^2\right)/\Im(\tau)} \chi^{\rm O }. 
$$
Since the additional factor is invariant under $\tau\to \tau+1$, we have full modular invariance:
\beq
\chi_{{\rm modular }}^{\rm O }(\tau\,|\vec v,\, \vec \xi)=
\chi_{{\rm modular }}^{\rm O }
\left({a\tau+b\over c\tau+d }\,\right|\left. {\vec v\over c\tau+d},\, {\vec \xi\over
c\tau+d}\right)
\label{modinv}
\eeq
The internal symmetry group parameters $\vec \xi$ are also acted upon by the modular group.  
\subsection{The  case of the $E_8\otimes E_8$ heterotic string}
Now we have 
\beq
\chi_R(\tau \,|\, \vec v,\,  \vec \xi)=e^{-2i\pi \tau} 
{\prod_{k} \sin\left(\pi v_k \right)\over \Theta^{(8)}_1(\vec v\,|\,\tau )}
\chi^{{\rm E }}(\tau \,|\, \vec \xi), 
\label{chiR8fac}
\eeq
where the last term is a $E_8\otimes E_8$ character. 
It is easy to see that  the Cartan 
algebras of $SO(32)$ and $E_8\otimes E_8$  are the same\footnote{The situation is similar to what
happens  between $O(2N)$ and $O(2N+1)$}. 
Thus the characters of the two string theories are computed from the 
same trace, albeit with different GSO like projections. 
There are now four sectors for the right movers. Since the $16$ dimensional space is 
split in two, we introduce the notation $\vec \xi \to \underline \xi, \underline \xi'$, 
where $\underline \xi=\{\xi_1, \ldots, \xi_{8}\}$, $\underline \xi'=\{\xi_{9}, \ldots, \xi_{16}\}$.
The result will be expressed as products of characters of $O(16)$. A calculation similar to the
one just summarised for the previous case gives  the internal symmetry contribution  
$$
\chi^{{\rm E }}(\tau \,|\, \vec \xi)=
$$
$$
=
{f_0^{-16}\over 4}
\left[\Theta^{(16)}_2(\underline\xi\, |\, \tau)+\Theta^{(16)}_1(\underline\xi\, |\, \tau)+
\Theta^{(16)}_3(\underline\xi\, |\, \tau)+\Theta^{(16)}_4(\underline\xi\, |\, \tau)\right]\times
$$
$$
\left[\Theta^{(16)}_2(\underline\xi'\, |\, \tau)+\Theta^{(16)}_1(\underline\xi'\, |\, \tau)+
\Theta^{(16)}_3(\underline\xi'\, |\, \tau)+\Theta^{(16)}_4(\underline\xi'\, |\, \tau)\right]
$$
According to Eq.\ref{Theta2Npm},  each term is a sum over 
the lattice $\Lambda_W^{{(16)e,h}}\oplus \Lambda_W^{{(16)e,i}}$  which is usually denoted 
 $\Gamma^8$. 
Putting together left and right movers  one finds 
$$
\chi^{\rm E }(\tau\,|\vec v,\, \vec \xi)=
\left({\Theta^{(8)}_1(\vec y\,|\,\tau)\over\Theta^{(8)}_1(\vec v \,|\,\tau)}\right)^*
{\left(\theta'(0|\tau)\right)^{4/3}  \over\Theta^{(8)}_1(\vec v \,|\,\tau)}\times 
$$
$$
{1\over 2}\left(\theta'(0|\tau)\right)^{-8/3}
\left[\Theta^{(16)}_2(\underline\xi\, |\, \tau)+\Theta^{(16)}_1(\underline\xi\, |\, \tau)+
\Theta^{(16)}_3(\underline\xi\, |\, \tau)+\Theta^{(16)}_4(\underline\xi\, |\, \tau)\right]\times
$$
\beq
{1\over 2}\left(\theta'(0|\tau)\right)^{-8/3} 
\left[\Theta^{(16)}_2(\underline\xi'\, |\, \tau)+\Theta^{(16)}_1(\underline\xi'\, |\, \tau)+
\Theta^{(16)}_3(\underline\xi'\, |\, \tau)+\Theta^{(16)}_4(\underline\xi'\, |\, \tau)\right]
\label{chE}
\eeq
Modular invariance is discussed in the same way as before, with the same additional factor as in 
Eq.\ref{modinv}. 
\section{The Narain lattices of compactification }
As is well known, The two heterotic string become T dual when compactified to nine dimensions. In
this situation the relevent zero mode lattices are made up by 
  appending  to either  $\Gamma^{16}$ or
$\Gamma^8\otimes \Gamma^8$ the even two dimensional  Lorentzian lattice $\Pi^{1,1}$.  
The mathematical tool  behind the T duality is that this  embedds them both into  the same  even
Lorentzian self-dual lattice 
$\Pi^{17,\, 1}$. The pratical realisation of this fact, which we will use, was put forward in
ref.\cite{G}.  Starting from the Euclidean self-dual lattices just
reviewed, the two embeddings   are most directly described using  two different sets of 
 basis vectors. The one associated with
$\Gamma^{(16)}$ will be described by    
$\vec e^{(16)}_i$, $i=1,\ldots, 16$,  $\vec k^{(16)}$, $\vec{\bar k}^{(16)}$, 
where the  first sixteen are unit vectors, and the last two are  null vectors 
orthogonal to the $e$'s
such that $\vec k^{(16)}.\vec{\bar k}^{(16)}=1$. The one associated with $\Gamma^8\otimes
\Gamma^8$ is similar; it  will be distinguished by a superscript $(8)$. The transformation between
the two basis will mix light-like vectors and Euclidean vectors. 

Group theoretically, the light-like vectors correspond to extensions of the root systems either of
$D_{16}$, or $E_8\otimes E_8$. Let recall the principle of these extensions for completeness,
following ref.\cite{GO}. Consider in general an  ordinary  simply laced Lie algebra,
 with roots $\vec \alpha_i$, $i=1,\ldots N$. The {\em  extended} root system is defined by
introducing  the additional root $\vec \alpha_0=-\sum_i n_i \vec \alpha_i+\vec k$, where $\vec
k^2=0$, 
$\vec \alpha_i\vec k=0$, and  $\sum_i n_i \vec \alpha_i$ is the biggest root. 
The {\em  over extended} root system involves yet another root 
$\vec \alpha_{-1}=-\left(\vec k+\vec {\bar k} \right)$,  with  another light-like vector  
$\vec {\bar k}$, such   $\vec k.\vec {\bar k}=1 $.  

 \subsection{Embedding of $\Gamma^{16}$ into $\Pi^{17,\, 1}$:}
Given the orthonormal set of $\vec e^{(16)}_i $ we may construct  the simple roots of $D_{16}$ as  
$$
\vec \pi^{(16)}_i=\vec e^{(16)}_i-\vec e^{(16)}_{i+1}, \quad i=1,\ldots, 15, \quad 
\vec \pi^{(16)}_{16}=\vec e^{(16)}_{15}+\vec e^{(16)}_{16}
$$
Applying, the procedure just recalled gives  the extensions
$$
\vec \pi^{(16)}_0= 
-(\vec e^{(16)}_1+\vec e^{(16)}_2)+\vec k^{(16)} , \quad 
\vec \pi^{(16)}_{-1}=-(\vec k^{(16)}+\vec{\bar k}^{(16)})
$$
In order to arrive at  the  $\Pi^{17,1}$ Dynkin 
diagram,   we introduce another vector 
$$
\vec \pi^{(16)}_{17}=-\demi\sum_{i=1}^{16} \vec e^{(16)}_i+\vec k^{(16)}-\vec{\bar
k}^{(16)}
$$
Denote the Cartan matrix  of $\Pi^{17,1}$ by  ${\cal K}_{\mu \nu}=\vec \pi^{(16)}_\mu . \vec
\pi^{(16)}_\nu$, with 
$\mu=-1,0,1,\ldots 17$. Then, apart from ${\cal K}_{\mu \mu}=2$,  the non zero entries above the
diagonal  are  
$$
{\cal K}_{ii+1}=-1, \quad i=1\ldots 14,\quad {\cal K}_{14, 16}=-1  
$$
which, of course,  give the diagram of $D_{16}$: and 
 $$
{\cal K}_{0, 2}=-1, \quad  {\cal K}_{-1, 0}=-1,\quad  {\cal K}_{17, 16}=-1. 
$$
 The Dynkin diagram is depicted on the following figure, where the original $D_{16}$ points are in
white\footnote{If you have a couloured picture, the  extended root is in blue, and the overextended
ones in red.}.
$$
\includegraphics{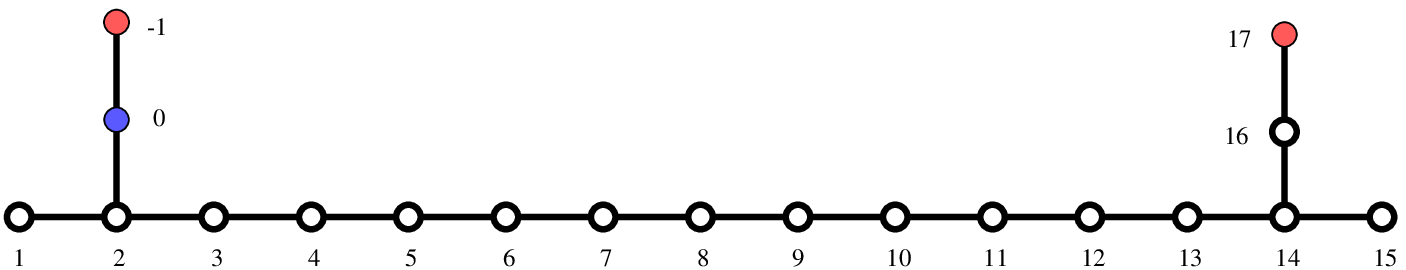}
$$    
The vectors
$\vec
\pi^{(16)}_i$,
$i=-1,0, 1,\ldots, 16$, are  
$18$ linearly independent   vectors, so 
that all others may be re-expressed in terms of them. 
Thus we may invert the transformation from the basis vectors to these  root vectors. 
One finds 
$$
\vec e^{(16)}_{16}=\demi\left(\vec \pi^{(16)}_{16}-\vec \pi^{(16)}_{15}\right),\quad 
\vec e^{(16)}_{15}=\demi\left(\vec \pi^{(16)}_{16}+\vec \pi^{(16)}_{15}\right),
$$
$$
\vec e^{(16)}_i=\sum_{j=i}^{14} \vec \pi^{(16)}_j+\demi\left(\vec \pi^{(16)}_{16}+\vec
\pi^{(15)}_{15}\right),\quad i=1,\ldots, 14.  
$$ 
\beq
\vec k^{(16)}=\vec \pi^{(16)}_0+\sum_{i=1}^{16}n^{(16)}_i \vec \pi^{(16)}_i,  
\quad 
\vec{\bar k}^{(16)}=-\vec \pi^{(16)}_{-1}-\vec \pi^{(16)}_0-\sum_{i=1}^{16}n^{(16)}_i \vec
\pi^{(16)}_i
\label{inv16}
\eeq  
Moreover $\vec \pi^{(16)}_{17}$ is not an independent vector. An explicit computation indeed shows
that 
\beq
 \vec \pi^{(16)}_{-1}+2\vec \pi^{(16)}_0+{3\over 2}\vec \pi^{(16)}_1+
\sum_{j=2}^{14}\left(4-{j\over
2}\right) \vec
\pi^{(16)}_j-{3\over 2}\vec\pi^{(15)}_{15}-2\vec \pi^{(16)}_{16} 
-\vec \pi^{(16)}_{17}
=0
\label{relat32}
\eeq
\subsection{Embedding of $\Gamma^8\otimes \Gamma^8$into $\Pi^{17,\, 1}$: }
We denote by $\vec \pi^{(1,8)}_i$, $\vec \pi^{(2,8)}_j$, $i,j=-1, 0, 1\ldots 8$ the two sets of
hyperextended simple roots of $E_8$. According to ref. \cite{G}   the same $\Pi^{17,\, 1}$ Dynkin
diagrams comes out from two hyperextended  $E_8$ Dynkin diagram glued together  if we use  the 
correspondence
$$
\vec \pi^{(1,8)}_i\Leftrightarrow\vec \pi^{(16)}_i,\> i=1,\ldots 6,\> 
 \vec \pi^{(1,8)}_7\Leftrightarrow\vec \pi^{(16)}_0, \>
 \vec \pi^{(1,8)}_8\Leftrightarrow\vec \pi^{(16)}_{-1}
$$
$$
\vec \pi^{(2,8)}_i\Leftrightarrow\vec \pi^{(16)}_{16-i},\> i=1,\ldots 6,\> 
 \vec \pi^{(2,8)}_7\Leftrightarrow\vec \pi^{(16)}_{16}, \>
 \vec \pi^{(2,8)}_8\Leftrightarrow\vec \pi^{(16)}_{17}
$$
\beq
\vec \pi^{(1,8)}_0\Leftrightarrow\vec \pi^{(16)}_{7},\> 
 \vec \pi^{(2,8)}_0\Leftrightarrow\vec \pi^{(16)}_{9}, \>
 \vec \pi^{(8)}_{-1}\Leftrightarrow\vec \pi^{(16)}_{8}
\label{8-16}
\eeq
The two diagrams have the $-1$ point in common. Thus we write $\vec \pi^{(8)}_{-1}$ instead of 
$\vec \pi^{(1,8)}_{-1}$ or $\vec \pi^{(2,8)}_{-1}$. This realisation is depicted in the following
picture with the same (colour) conventions as the preceding one. 
$$
\includegraphics{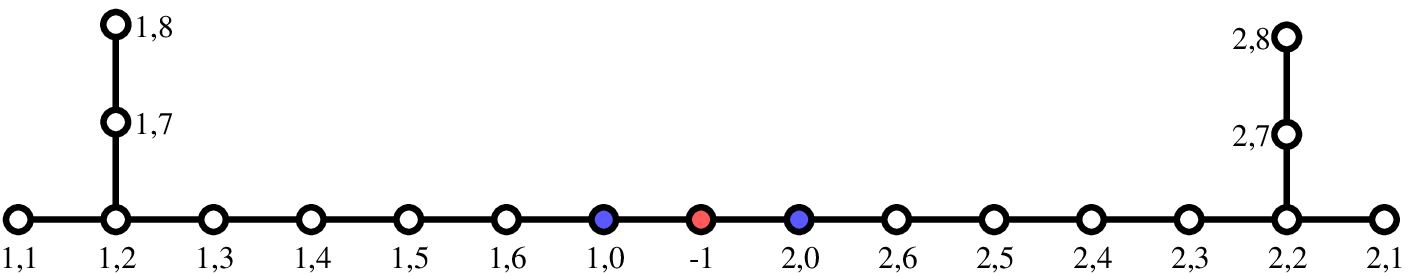} 
$$ 
We make use of the explicit
realisation of ref.\cite{G}: 
$$
\vec \pi^{(1,8)}_i=\vec e^{(8)}_i-\vec e^{(8)}_{i+1},\> i=1,\ldots, 6,\> 
\vec \pi^{(1,8)}_7=-\vec e^{(8)}_1-\vec e^{(8)}_{2},\> 
\vec \pi^{(1,8)}_8=\demi \sum_{i=1}^8 \vec e^{(8)}_i
$$
$$
\vec \pi^{(2,8)}_i=\vec e^{(8)}_{16-i}-\vec e^{(8)}_{17-i},\> i=1,\ldots, 6,\>
\vec \pi^{(2,8)}_7=\vec e^{(8)}_{15}+\vec e^{(8)}_{16},\> 
\vec \pi^{(2,8)}_8=-\demi \sum_{i=9}^{16} \vec e^{(8)}_i
$$
$$
\vec \pi^{(1,8)}_0= \vec e^{(8)}_7-\vec e^{(8)}_8+\vec k^{(8)},\quad 
\vec \pi^{(2,8)}_0=\vec e^{(8)}_9-\vec e^{(8)}_{10}+\vec k^{(8)}, 
\quad \vec \pi^{(8)}_{-1}=-\vec k^{(8)}-{\vec{\overline k}}^{(8)}
$$
Then one may verify that this set of vectors does generate the same $\Pi^{17,1}$ Cartan matrix. 
 Concerning the linear dependence, one may check that the $(8)$ root vectors
satisfy, for $\ell=1,2$,   
\beq
\vec \pi^{(\ell,8)}_{8}+2\vec \pi^{(\ell,8)}_{7}+{3\over 2} \vec \pi^{(\ell,8)}_{1}+
\sum_{j=2}^6 (4-{j\over 2})\vec \pi^{(\ell,8)}_{j}+{1\over 2}\vec \pi^{(\ell,8)}_{0}
={1\over 2}\vec k^{(8)}, 
\label{null1}
\eeq
so that  
$$
\vec \pi^{(1,8)}_{8}+2\vec \pi^{(1,8)}_{7}+{3\over 2} \vec \pi^{(1,8)}_{1}+
\sum_{j=2}^6 (4-{j\over 2})\vec \pi^{(1,8)}_{j}+{1\over 2}\vec \pi^{(1,8)}_{0}
$$
\beq
-\vec \pi^{(2,8)}_{8}-2\vec \pi^{(2,8)}_{7}-{3\over 2} \vec \pi^{(2,8)}_{1}-
\sum_{j=2}^6 (4-{j\over 2})\vec \pi^{(2,8)}_{j}-{1\over 2}\vec \pi^{(2,8)}_{0}=0
\label{relat8}
\eeq
This coincides with the transformed of Eq.\ref{relat32} through the correspondence \ref{8-16}
\subsection{The transformation between basis vectors:}
Since the linear dependence of the extended root vectors is the same, we may look for a
transformation between the basis vectors, such that the root vectors coincide following 
the correspondence displayed on Eq.\ref{8-16}. Making use of Eq.\ref{inv16}, one finds
\beq
\vec e^{(16)}_{i} = \vec e^{(8)}_{i},\quad i=10, \ldots 16 
\label{trbasis10-16}
\eeq
\beq 
\vec e^{(16)}_9 = \vec e^{(8)}_9+\vec k^{(8)}, \quad 
\vec e^{(16)}_8-\demi \vec k^{(16)}=\vec e^{(8)}_8-k^{(8)} 
\label{trbasis8-9}
\eeq 
\beq
\vec e^{(16)}_{i}-\demi \vec k^{(16)} = \vec e^{(8)}_{i},\quad i=1, \ldots 7 
\label{trbasis1-7}
\eeq
\beq
\vec k^{(16)}=2\left(\vec e^{(8)}_9-\vec e^{(8)}_8+\vec k^{(8)}-{\vec{\overline k}}^{(8)}\right)
\label{trbasis0}
\eeq
\beq
\vec{\bar k}^{(16)}= -\demi \sum_{i=1}^8 \vec e^{(8)}_i
-2\left(\vec e^{(8)}_9-\vec e^{(8)}_8+\vec k^{(8)}-{\vec{\overline k}}^{(8)}\right)
\label{trbasis0b}
\eeq
For future discussion, let us consider the transformation of a generic lattice point. 
Imposing that  
$$
\sum_{\mu=1}^{16} m^{(16)}_\mu \vec e^{(16)}_\mu+m^{(16)}_0\vec k^{(16)}
+ \bar m^{(16)}_0  \vec{\bar k}^{(16)}=\sum_{\mu=1}^{16} m^{(8)}_\mu \vec e^{(8)}_\mu+m^{(8)}_0\vec
k^{(8)} + \bar m^{(8)}_0  \vec{\bar k}^{(8)}
$$
gives 
 \beq
m^{(8)}_\mu= m^{(16)}_\mu-\demi \bar m^{(16)}_0,\quad \mu=1,\ldots, 7, 
\label{tr1-7}
\eeq
 \beq
m^{(8)}_8=-\sum_{\mu=1 }^{7} m^{(16)}_\mu-2m^{(16)}_0+{3\over 2}\bar m^{(16)}_0  
\label{tr8}
\eeq
 \beq
m^{(8)}_9=\sum_{\mu=1 }^{7} m^{(16)}_\mu+m^{(16)}_8+m^{(16)}_9+2m^{(16)}_0-2\bar m^{(16)}_0
\label{tr9}
\eeq
\beq
 m^{(8)}_\mu=m^{(16)}_\mu,\quad \mu=10,\ldots, 16, 
\label{tr10-16}
\eeq
\beq
m^{(8)}_0=\sum_{\mu=1 }^{7} m^{(16)}_\mu+m^{(16)}_9+2m^{(16)}_0-2\bar m^{(16)}_0
\label{tr0}
\eeq
\beq
\bar m^{(8)}_0=-\sum_{\mu=1 }^{7} m^{(16)}_\mu-m^{(16)}_8-2m^{(16)}_0+2\bar m^{(16)}_0
\label{tr0bar}
\eeq
Note, in particular that  
\beq
\sum_{\mu=1}^8 m^{(8)}_\mu= -2\left(m^{(16)}_0+\bar m^{(16)}_0\right)  
\label{sum8}
\eeq
\beq
\sum_{\mu=9}^{16} m^{(8)}_\mu=\sum_{\mu=1 }^{16} m^{(16)}_\mu+2m^{(16)}_0-2\bar
m^{(16)}_0
\label{sum9}
\eeq
One thus sees that, if $m^{(16)}_0$, and $\bar m^{(16)}_0$ are integer, and $\sum_{\mu=1 }^{16}
m^{(16)}_\mu$ is even, $\sum_{\mu=1}^8 m^{(8)}_\mu$, and $\sum_{\mu=9}^{16} m^{(8)}_\mu$ are 
also both even. 
Moreover, it follows from the above that $m^{(8)}_0$, and $\bar m^{(8)}_0$ are both integer. 
Clearly, by  construction, we have the orthogonality relations
$$
\sum_{\mu=1}^{16} \left(m^{(16)}_\mu\right)^2+2 m^{(16)}_0 \bar m^{(16)}_0=
\sum_{\mu=1}^{16} \left(m^{(8)}_\mu\right)^2+2 m^{(8)}_0 \bar m^{(8)}_0. 
$$
Let us denote by $\Gamma^{(1,1)}$ the set of points $p\vec k+q\vec{\bar k}$, $p,q\in {\cal Z}$. 
Then  Eqs.\ref{tr1-7}--\ref{tr0bar} define  a $SO(17,1)$ transformation such that 
$\Gamma^8\otimes \Gamma^8\otimes \Gamma^{(1,1)}\Leftrightarrow \Gamma^{16}\otimes \Gamma^{(1,1)}$.
\subsection{Intermediate basis vectors:}
Quite generally, given  a set of basis vectors $\vec e_i$, $i=1,\ldots 16$, $\vec k$ and $\vec {\bar
k}$, we  get another set satisfying the same orthogonality relations by letting
$$
{\vec e}'_i= \vec e_i-\beta_i \vec k,\quad 
\vec {\bar k} ' = \sum_j \beta_j \vec e_j+  \vec {\bar k}  -\demi \sum_j\beta_j^2 \vec k, \quad 
\vec k'=\vec k. 
$$
where $\beta_i$ are arbitrary. Looking at the transformation formulae \ref{trbasis10-16} -- 
\ref{trbasis0b} , one
sees that such transformations do appear on each side with  
\beq
\beta^{(16)}_{i}= \demi,\>  {\rm  for }\> i=1,\ldots 8,\quad 
\beta^{(16)}_{i}= 0,\> {\rm  for }\> i=9,\ldots 16
\label{beta16}
\eeq
\beq
\beta^{(8)}_{8}=-\beta^{(8)}_{9} =1,\quad 
\beta^{(16)}_{i}= 0, \> {\rm  for }\> i\not= 9, {\rm or  }\> 8
\label{beta8}
\eeq
Let us  introduce, the corresponding basis vectors with a prime. Eqs.\ref{trbasis10-16} --
\ref{trbasis0b} leads to  
\beq
\vec e^{'(16)}_{i} = \vec e^{'(8)}_{i},\quad i=1, \ldots 16,\quad 
\vec k^{'(16)}=-2 \vec {\bar k'}^{(8)}, \quad 
 \vec {\bar k'}^{(16)}=-\demi  \vec k^{'(8)}.  
\label{pmtrs}
\eeq
With these intermediate bases, the transformation reduces to  simple exchange and rescaling of the
light-like vectors.   

Next, it is easy  to determine the range of the transformed coordinates also
distinguished by a prime. We will denote by $\Gamma^{(17,1)'}_{(16)}$ and $\Gamma^{(17,1)'}_{(8)}$ 
the corresponding lattices. They are best visualised using the   $SO(16)\otimes SO(16)$ weight
lattices obtained by separating the $m_\mu'$ variables with $\mu\not=0$ into  two sets made 
up with  the eight
first and with  the  eight last variables, respectively. One finds\footnote{  
${\cal Z}$ denotes the set of non negative integers.} 
\begin{center}
\begin{tabular}{||cc||c||}\hline\hline  
$m_0^{(16)'}$ & $\bar m_0^{(16)'}$&$D_8\otimes D_8$   weight space\\
\hline  
$ {\cal Z} $      & even & 
                          $\left(\Lambda^{(8)ei}\otimes  \Lambda^{(8)ei}\right)\oplus 
                           \left(\Lambda^{(8)eh}\otimes \Lambda^{(8)eh}\right)$  \\
${\cal Z} $       & odd  &  $\left(\Lambda^{(8)ei}\otimes \Lambda^{(8)eh}\right)\oplus  
                            \left(\Lambda^{(8)eh}\otimes \Lambda^{(8)ei}\right)$\\
${\cal Z} +\demi$ & even &  $\left(\Lambda^{(8)oi}\otimes \Lambda^{(8)oi}\right) \oplus 
                            \left(\Lambda^{(8)oh}\otimes \Lambda^{(8)oh}\right)$ \\
${\cal Z}+\demi$  & odd  &  $\left(\Lambda^{(8)oi}\otimes \Lambda^{(8)oh}\right) \oplus 
                             \left(\Lambda^{(8)oh}\otimes \Lambda^{(8)oi}\right)$ \\
\hline\hline 
\end{tabular}
\end{center}
\begin{center}
\begin{tabular}{||cc||c||}\hline\hline  
$m_0^{(8)'}$ & $\bar m_0^{(8)'}$&$D_8\otimes D_8$   weight space  \\
\hline  
$ {\cal Z} $      & even & 
                          $\left(\Lambda^{(8)ei}\otimes  \Lambda^{(8)ei}\right)\oplus 
                           \left(\Lambda^{(8)eh}\otimes \Lambda^{(8)eh}\right)$  \\
${\cal Z} $       & odd  &  $\left(\Lambda^{(8)oi}\otimes \Lambda^{(8)oh}\right)\oplus  
                            \left(\Lambda^{(8)oh}\otimes \Lambda^{(8)oi}\right)$\\
${\cal Z} +\demi$ & even &  $\left(\Lambda^{(8)ei}\otimes \Lambda^{(8)ei}\right) \oplus 
                            \left(\Lambda^{(8)eh}\otimes \Lambda^{(8)eh}\right)$ \\
${\cal Z}+\demi$  & odd  &  $\left(\Lambda^{(8)oi}\otimes \Lambda^{(8)oh}\right) \oplus 
                             \left(\Lambda^{(8)oh}\otimes \Lambda^{(8)oi}\right)$ \\
\hline\hline 
\end{tabular}
\end{center} 

Let us turn to  the
transformation between the
$m$'s. One verifies that
\beq
m^{'(16)}_\mu =m^{'(8)}_\mu,\quad   m^{'(16)}_0=-\demi \bar m^{'(8)}_0,\quad 
\bar m^{'(16)}_0=-2  m^{'(8)}_0
\label{prelat2}
\eeq
This is indeed a mapping  $\Gamma^{(17,1)'}_{(16)}\leftrightarrow \Gamma^{(17,1)'}_{(8)}$. It
corresponds to 
\begin{center}
\begin{tabular}{||cc||cc||}\hline\hline  
$m_0^{(16)'}$ & $\bar m_0^{(16)'}$&$m_0^{(8)'}$ & $\bar m_0^{(8)'}$  \\
\hline  
$ {\cal Z} $      & even    &$ {\cal Z}  $ &  even  \\
${\cal Z} $       & odd     & $ {\cal Z}+\demi $ & even   \\
${\cal Z} +\demi$ & even    & $ {\cal Z} $ &   odd \\
${\cal Z}+\demi$  & odd     & $ {\cal Z}+\demi $ & odd   \\
\hline\hline 
\end{tabular}
\end{center}
The two lines in the middle are interchanged.
\section{Compactified heterotic characters}
\subsection{Definition}
The characters of course involve summations over the discrete momentum and winding number in a way 
similar to the purely bosonic and type II cases,  described in ref.\cite{CG}. The left movers, and
the space-time part of the right movers  contributions are straightforwardly written down, following
the line of this  last reference. Thus  Eq.\ref{ch32} are  replaced by  equations of the form   
$$
\chi^{\rm O }(\tau, \,R\,  |\vec v^{(3)},\, \vec \xi)=
\left(\left(\theta'(0|\tau)\right)^{2/3}{\Theta^{(8)}_1(\vec 
y^{(3)}\,|\,\tau)\over\Theta^{'(8)}_1(\vec v^{(3)}
\,|\,\tau)}\right)^* {\left(\theta'(0|\tau)\right)^{4/3}  \over\Theta^{'(8)}_1(\vec v^{(3)}
\,|\,\tau)}\chi_{(16)}(\tau\, R\, |\, \vec \xi),  
$$
with the same  form  for   Eq.\ref{chE} (for $\chi^{\rm E }$) involving  
$\chi_{(8)}(\tau\, R\, |\, \vec \xi)$ instead of  $\chi_{(16)}(\tau\, R\, |\, \vec \xi)$. 
For the following it is convenient to write   both formulae at once   with an index
$\ell=16$ for $SO(32)$, or $8$  for $E_8\otimes E_8$, respectively. We have pulled out 
explicit modular
invariant factors, so that when we come to it, we will only have to check the modular 
invariance  of $\chi_{(\ell)}$. 
In the above formulae, we  have gone from  $8$ to $7$ transverse dimensions. Choosing to
compactify the eighth direction, the corresponding breaking of rotational invariance forces us to
let  
$v_4=0$. We write  $\vec v^{(3)}=\{ v_1,\, v_2,\, v_3, 0\}$, and note  $\vec y^{(3)}$, 
its transformed by triality, that is 
$$
y^{(3)}_1=\demi( v_1-v_2+v_3),\quad  y^{(3)}_2=\demi(-v_1+v_2+v_3),
$$
$$
y^{(3)}_3=\demi( v_1+v_2+v_3),\quad  y^{(3)}_4=\demi(- v_1-v_2+v_3),
$$ 
Using the same convention as in  ref.\cite{CG}, we have let
$\Theta^{'(8)}_1(\vec v^{(3)}|\tau)\equiv  \theta'_1(0|\tau)\prod_{\mu=1}^3 \theta_1(v_\mu|\tau)$. 

Let us now turn to  the characters $\chi_{(\ell)}$ that involve the summation over internal symmetry
and compactified mode  lattices.  Their definition is most natural using the lattices
$\Gamma^{(17,1)'}_{(\ell)}$. Indeed, the corresponding theories are explicitly invariant under
$SO(16)\otimes SO(16)$, so that we may use the standard mathematical definition of characters which
is  based on the corresponding $16$ dimensional Cartan algebra. Moreover, with this lattice, we have
seen that   
 the T duality transformations does not involve the internal symmetry modes. Duality will thus
be very simply ensured.    With these motivations, we introduce  the  definitions  
$$
\chi_{(\ell)}(\tau |\, R,\, \vec \xi )={1\over \sqrt{\tau-\tau^*}}
\left(\left(\theta'_1(0|\tau)\right)^*\right)^{-2/3}
\left(\theta'_1(0|\tau)\right)^{-18/3}\times  
$$
\beq
\sum_{ \Gamma_{(\ell)}^{(17,1)'} }
e^{i\pi[\tau(m'{}^{(\ell)}_0/2R+\bar m'{}^{(\ell)}_0R)^2-
\tau^*(m'{}^{(\ell)}_0/2R-\bar m'{}^{(\ell)}_0R)^2]}  
 e^{i\pi \tau \sum_{\mu=1}^{16} \left(m'{}^{(\ell)}_\mu\right)^2} 
e^{2i\pi\sum_{\mu=1}^{16} m'{}^{(\ell)}_\mu\xi_\mu }
\label{Spsldef}
\eeq
\subsection{Duality}
 Let us use the mapping beween the two lattices spelled out on Eqs.\ref{tr1-7} -- \ref{tr0bar}. 
One gets after the corresponding change of variable, 
$$
\chi_{(16)}(\tau |\, R,\, \vec \xi )=
{1\over \sqrt{\tau-\tau^*}}
\left(\left(\theta'_1(0|\tau)\right)^*\right)^{-2/3}
\left(\theta'_1(0|\tau)\right)^{-18/3}\times  
$$
$$
\sum_{ \Gamma_{(8)}^{(17,1)'} }
e^{i\pi[\tau(2 m'{}^{(8)}_0R+ \bar m'{}^{(8)}_0/4R)^2-
\tau^*(2 m'{}^{(8)}_0R-\bar m'{}^{(8)}_0/4R)^2]}  
 e^{i\pi \tau \sum_{\mu=1}^{8} \left(m'{}^{(8)}_\mu\right)^2} 
e^{2i\pi\sum_{\mu=1}^{16} m'{}^{(8)}_\mu\xi_\mu }, 
$$
which gives 
$$
\chi_{(16)}(\tau |\, R,\, \vec \xi )=
\chi_{(8)}(\tau |\, 1/4R,\, \vec \xi ). 
$$
\subsection{Modular invariance}
Applying Eqs.\ref{Theta2Npm}, it is straightforward to re-express $\chi_{(\ell)}$ 
in terms of $\Theta^{(16)}_i$. The result is conveniently written under the form 
\beq
\chi_{(\ell)}(\tau |\, R,\, \vec \xi )=
\left(\theta'(0|\tau)\right)^{-16/3} \sum_{i, jk} N^{(\ell)}_{i, jk}F_i(R,\tau) 
\Theta_j^{(16)}(\underline \xi\, |\tau )\Theta_k^{(16)}(\underline \xi'\, |\tau ) 
\label{chiltheta}
\eeq 
where $\ell=16, 8 $, and    
\beq
F_i(R,\tau) ={1\over \sqrt{\tau-\tau^*}}
\left|{1\over \left(\theta'_1(0|\tau)\right)^{2/3}}\right|^2
\sum_{m ,n\in \Lambda^{(2)}_i } e^{i\pi (\tau-\tau^*)[m^2/4R^2+
n^2R^2]}e^{-i\pi(\tau+\tau^*)mn]}. 
\label{Fidef}
\eeq
The  lattices $\Lambda^{(2)}_i$, $i=1,\ldots, 4$  are two dimensional lattices of $R^2$, with points
of coordinates ($m,\, n$) that are  (integer, even), (integer, odd), 
(half integer, even), (half integer, odd) respectively. 
 The numerical coefficients $ N^{(\ell)}_{i, jk}$, considered as matrices in the
last two indices  are defined as follows. 
$$
N^{(16)}_1=\left(\begin{array}{cccc}
1&1&0&0\\
1&1&0&0\\
0&0&1&1\\
0&0&1&1
\end{array}\right),\quad 
N^{(16)}_2=\left(\begin{array}{cccc}
0&0&1&1\\
0&0&1&1\\
1&1&0&0\\
1&1&0&0
\end{array}\right)
$$
$$
N^{(16)}_3=\left(\begin{array}{cccc}
1&-1&0&0\\
-1&1&0&0\\
0&0&1&-1\\
0&0&-1&1
\end{array}\right),\quad 
N^{(16)}_4=\left(\begin{array}{cccc}
0&0&-1&1\\
0&0&1&-1\\
-1&1&0&0\\
1&-1&0&0
\end{array}\right)
$$
$$
N^{(8)}_1=N^{(16)}_1,\quad  N^{(8)}_2=N^{(16)}_3, 
\quad  N^{(8)}_3=N^{(16)}_2, \quad  N^{(8)}_4=N^{(16)}_4 
$$
\subsubsection{The transformation $\tau\to -1/\tau $} 
The behaviour of the Theta functions is summarised by the matrix $S$ of Eq.\ref{Sdef}. After some
computation, one verifies that, for both $\ell=16,\, 8$,   
\beq
N^{(\ell)}_k=\sum_\ell {\cal S }_{kj}S N^{(\ell)}_j S,\qquad 
{\cal S}=\demi \left(\begin{array}{cccc}
1&1&1&1\\
1&1&-1&-1\\
1&-1&1&-1\\
1&-1&-1&1
\end{array}\right)
\label{mod16}
\eeq
Concerning the functions $F_i$, They all may be deduced from the twisted characters. 
$$
F(R,\tau, \vec \epsilon, \vec y)={1\over \sqrt{\tau-\tau^*}}
\left|{1\over \left(\theta'_1(0|\tau)\right)^{2/3}}\right|^2\times 
$$
$$
\sum_{m ,n\in {\cal Z } } e^{i\pi (\tau-\tau^*)[(m+\epsilon_1/2)^2/4R^2+
(n+\epsilon_2/2)^2R^2]}e^{-i\pi(\tau+\tau^*)(m+\epsilon/2)(n+\epsilon_2/2)]}e^{i\pi(\eta_1m+ \eta_2
n)}, 
$$
where $\vec \epsilon$, $\vec \eta$ have components one or zero. Applying Poisson resummation
technics, to this twisted character, one verifies that the functions $F_i$ satisfy
\beq
\underline F(R,\tau)=
 {\cal S }\underline F(R,-1/\tau),\quad{\rm where\>  } 
\underline F \equiv \left(\begin{array}{c}
F_1\\
F_2\\
F_3\\
F_4
\end{array}\right)
\label{Ftr}
\eeq
Finally, making use of Eqs.\ref{chiltheta}, \ref{mod16},  the last equation, and the fact that 
${\cal S }^2=1$ one verifies that  $\chi_{(\ell)}(\tau |\, R,\, \vec \xi )$ is invariant under 
$\tau\to -1/\tau$.
\subsubsection{The transformation $\tau\to \tau+1$}
The discussion is similar but simpler. One has 
$$
\Theta^{(2N)}_i(\vec\xi\, |\, \tau+1)=T_{ij}\Theta^{(2N)}_j(\vec\xi\, |\, \tau), \quad 
T=\left(\begin{array}{cccc}
1&0&0&0\\
0&1&0&0\\
0&0&0&1\\
0&0&1&0
\end{array}\right)
$$
$$
TN^{(\ell)}_iT={\cal T }_{ij}N^{(\ell)}_i,\quad  {\cal T }=\left(\begin{array}{cccc}
1&0&0&0\\
0&1&0&0\\
0&0&1&0\\
0&0&0&-1
\end{array}\right) 
$$
$$
\underline  F(\tau+1)={\cal T  } \underline F(\tau). 
$$
Modular invariance follows from ${\cal T  }^2=1$. 
\subsection{The decompactification limits }
In Eq.\ref{Fidef}, for $R\to \infty$ only $n=0$ may give a non vanishing contribution. 
Thus $F_2$, and
$F_4$ vanish for $R\to \infty$, and 
$F_1\sim F_3$. Thus one gets, taking  $\ell=16$ for definiteness, 
$$
\sum_{i, jk} N^{(16)}_{i, jk}F_i(R,\tau) 
\Theta_j^{(16)}(\underline \xi\, |\tau )\Theta_k^{(16)}(\underline \xi'\, |\tau )
$$
$$
\to_{R\to \infty} F_1(\infty, \tau) \sum_{ jk} \left(N^{(16)}_{1, jk}+N^{(16)}_{3,
jk}\right) 
\Theta_j^{(16)}(\underline \xi\, |\tau )\Theta_k^{(16)}(\underline \xi'\, |\tau )
$$
$$
= F_1(\infty, \tau)  
\sum_{i} 
\Theta_i^{(16)}(\underline \xi\, |\tau )\Theta_i^{(16)}(\underline \xi'\, |\tau )=
F_1(\infty, \tau) \sum_{i} 
\Theta_i^{(32)}(\vec \xi\, |\tau ). 
$$
Indeed this gives back the lattice $\Gamma^{(16)}$. 
on the other hand,  for $R\to 0$, only $m=0$ may give a non vanishing contribution. Thus $F_3$, and
$F_4$ disappear, and 
$F_1\sim F_2$. Thus one gets,
 $$
\sum_{i, jk} N^{(16)}_{i, jk}F_i(R,\tau) 
\Theta_j^{(16)}(\underline \xi\, |\tau )\Theta_k^{(16)}(\underline \xi'\, |\tau )
$$
$$
\to_{R\to 0} F_1(0, \tau) \sum_{ jk} \left(N^{(16)}_{1, jk}+N^{(16)}_{2,
jk}\right) 
\Theta_j^{(16)}(\underline \xi\, |\tau )\Theta_k^{(16)}(\underline \xi'\, |\tau )
$$
$$
= F_1(\infty, \tau)  
\sum_{i,j } 
\Theta_i^{(16)}(\underline \xi\, |\tau )\Theta_j^{(16)}(\underline \xi'\, |\tau )=. 
$$
This gives back the character associated with the  lattice $\Gamma^{(8)}\otimes \Gamma^{(8)}$.  

One sees that the characters we have introduced define a one parameter set of theories which
interpolate between the uncompactified $SO(32)$ heterotic string (obtained for $R\to \infty$) and
the $E_8\otimes E_8$ one (obtained for $R\to 0$). These theories should be considered as obtained by
a sort of twisted compactification, where a non trivial ``coupling'' is introduced between the various
sectors of the compactified modes (the functions $F_i$ and the $SO(16)$ modular blocks $\Theta_i$. 
\subsection{Return to the self-dual lattices.} 
 Finally, we consider the theories compactified in the usual way, using the lattices 
$\Gamma^{16}$ and
$\Gamma^8\oplus \Gamma^8$. The way to ensure T duality is to define the characters as equal to
the ones just introduced, and to retransform back the lattice summation variables 
to the variables $m_\mu ^{(\ell)}$ using the formulae displayed in section 3.   One finds 
$$
\chi _{(\ell)}(\tau|\, R,\, \vec \xi )={1\over \sqrt{\tau-\tau^*}}
\left(\left(\theta'_1(0|\tau)\right)^*\right)^{-2/3}
\left(\theta'_1(0|\tau)\right)^{-18/3}\times
$$
\beq  
\sum_{{m_0{(\ell)} ,\bar m_0{(\ell)} \in {\cal Z }\atop
\vec m{}^{(\ell)}\in \Gamma^{(\ell)}}} e^{i\pi[
\tau(p_R^{(\ell)})^2-\tau^*(p_L^{(\ell)})^2]}   
e^{2i\pi
\left[ \left(\vec m{}^{(\ell)}+\vec A{}^{(\ell)}\omega^{(\ell)}
\right)\vec \xi\right] }
\label{childef}
\eeq 
\beq
\left(p_R^{(\ell)}\right)^2=
\left\{\demi p^{(\ell)}-{1\over 4}  \left(\vec A^{(\ell)}\right)^2 \omega^{(\ell)}-
\demi \vec A^{(\ell)}\vec m^{(\ell)}
+\omega{}^{(\ell)}\right\}^2+ \left(\vec m^{(\ell)}+\vec A^{(\ell)}\omega{}^{(16)}\right)^2, 
\label{pRldef}
\eeq
\beq
\left(p_L^{(\ell)}\right)^2=
\left\{\demi p^{(\ell)}-{1\over 4}  \left(\vec A^{(\ell)}\right)^2 \omega^{(\ell)}-
\demi \vec A^{(\ell)}\vec m^{(\ell)}
-\omega{}^{(\ell)}\right\}^2, 
\label{pLldef}
\eeq
$$
p^{(\ell)}=m^{(\ell)}_0/R,\quad \omega^{(\ell)}=\bar m^{(\ell)}_0 R,
$$  
$$
A^{(16)}_\mu=-1/2R,\quad \mu=1,\ldots, 8,\quad A^{(16)}_\mu=0,\quad \mu=9,\ldots, 16, 
$$ 
$$
A^{(8)}_8=-1/R,\quad A^{(8)}_9=1/R,\quad A^{(8)}_\mu=0,\quad \mu\not=\, 8, 9.
$$ 
Of course  Wilson lines appear, and these formulae show how to define characters when non  zero
background gauge fields are turned on. Note the presence of  a coupling between the $\xi$ variables
and winding number: $\exp\left(2\pi i \left(\vec A^{(\ell)}\vec \xi \omega^{(\ell)}
\right)\right)$. 

At this point one should keep in mind that, although the characters are the same as in the
intermediate theories, the string theories are of course different, since the mass operators is not
invariant under $SO(17,1)$.    
\section{Outlook}
In the same way as for type II superstrings \cite{CG} the present study of Lie group characters
gives an interesting insight  at the perturbative level. The basic advantage, as compared to
partition functions, is that the characters we have studied encode the full structure of the
perturbative states. Moreover the  $SO(17,1)$ that relate
the weight lattices act naturally on them, giving a detailed relationship between perturbative
spectra. This revealed the existence of the  intermediate theories where T duality is simply the
inversion of the compactification radius. It is possible that this theory will play a useful role
in connecting the various string/M theories, since in that game the two heterotic strings
theories compactified on a circle  are usually not distinguished, owing to their exchange by 
T duality
although their structures are very different as we have seen.

We have studied the simplest compactification. Of course it would be interesting to sudy
compactifications on higher dimensional manifolds. One may expect that a similar discussion
could be carried out, but the needed mathematical properties are much more involved and much
less developed. 

So far we did not select out the BPS states. This may be done trivially, by simply retaining
only the zero mode contribution in $\Theta^{(8)}\left(\vec y|\tau\right)$ in Eq.\ref{chiL}. 
Then modular invariance is destroyed. Nevertheless, one might  try to extend the character
formulae (as well as the ones written in ref.\cite{CG}) to non perturbative states,  so that S duality
could be verified by change of variables in the summations that define them. This very strong
requirement would open the way towards understanding the characters of M theory.        

\noindent {\large \bfseries Acknowledgements}  This research  was partly carried out while I was
enjoying the warm hospitality, the very stimulating atmosphere, and the generous financial 
support of
the Newton Institute in Cambridge UK.  I am indebted to E. Cremmer for collaboration at
an early stage of this work. Discussions with C. Bachas,  D. Olive, and A. Schwimmer  have been very
helpful. 
\begin{appendix}
\section{The Heterotic String }
For completeness, I summarise here some elementary facts about heterotic strings in the  light
cone formulation.  Notations are the same as in \cite{GSW}, except that it is notationally
simpler to exchange left and right movers. Thus the  left movers are described by the same
operators   
$\widetilde \alpha_n^\mu$ ($\mu=1\ldots 8$),  and $S_n^a$  ($n\in {\cal Z}$, $a$ $O(8)$ spinor
index) as the type II superstrings. The mass operator is 
$$
\widetilde N=\sum_{n\geq 1}\left(  \widetilde\alpha^\mu _{-n}\widetilde \alpha^\mu_{n}
+ n\widetilde S^a_{-n}\widetilde S^a_{n}\right). 
$$
\subsection*{The  $SO(32)$ heterotic string revisited}  
For the right movers, we have the right-moving part of $X^\mu$ and $\lambda^A$, with $A=1,\ldots 32$
where the latter are world sheet fermions.  There are two sectors:
\paragraph{The (periodic) P sector:} There 
$$
\lambda^A(\sigma)=\sum_{{\rm \>integer}} e^{-in\sigma} \lambda_n^A,\quad 
\left[\lambda_n^A, \lambda_m^B\right]_+=\d(A,B)\d(n,-m).
$$
The right-right matching condition is 
$$
\widetilde N= N-1,\quad N=\sum_{n\geq 1}\left( \widetilde \alpha^i_{-n}\widetilde\alpha^i_{n}
+ n\lambda^A_{-n}\lambda^A_{n}\right). 
$$
 One defines 
$$
(-1)^F=\overline \lambda_0 (-1)^{\sum_{n\geq 1}\lambda^A_{-n}\lambda^A_{n}}
$$
where
$$
\overline \lambda_0 =\lambda^1_0 \cdots \lambda^{32}_0
$$
and only keeps the eigenvalue $+1$.    
\paragraph{The (antiperiodic) A sector:} There  
$$
\lambda^A(\sigma)=\sum_{r {\rm \>half \>integer}} e^{-ir\sigma} \lambda_n^A,\quad 
\left[\lambda_r^A, \lambda_s^A\right]_+=\d(A,B)\d(r,-s).
$$ 
The matching condition is 
$$
N=\widetilde N-1,\quad 
\widetilde N=\sum_{n\geq 1}\left( \widetilde \alpha^i_{-n}\widetilde \alpha^i_{n}
+ r\lambda^A_{-r}\lambda^A_{r}\right). 
$$
 One defines 
$$
(-1)^F= (-1)^{\sum_{r\geq 1/2}\lambda^A_{-r}\lambda^A_{r}}
$$
and only keeps the eigenvalue $+1$.
\subsection*{The $E_8\otimes E_8$ heterotic string}
The right sector is described by the same  oscillators with different boundary conditions for the
$\lambda^A$ operators. Separate 
$A=1,\,\ldots ,\, 16$, denoted $\lambda^A$ and $A=17,\,\ldots ,\, 32$ denoted $\lambda'^A$. 
One assignes  independently $A$ and $P$ boundary 
conditions to each set, and  GSO projectors $(-1)^{F_i}$, $i=1,2$ for each set. 
The physical states are taken to have eigenvalue $+1$ for both parity charges.  
The level matching conditions are
$$
\left\{ 
\begin{array}{cc}
\widetilde N=N-1&{\rm \>in \>the\>  }AA\> {\rm sector }\nonumber\\
\widetilde N=N&{\rm \>in \>the\>  }AP\>{\rm and }PA \> {\rm sectors }\nonumber\\
\widetilde N=N+1&{\rm \>in \>the\>  }PP\> {\rm sector }\nonumber\\
\end{array}\right.
$$
\end{appendix}

\fin